\documentclass[times,twocolumn,final]{elsarticle}

\usepackage{rinp}
\usepackage{framed,multirow}

\usepackage{amssymb}
\usepackage{latexsym}

\usepackage{url}
\usepackage{xcolor}
\definecolor{newcolor}{rgb}{.8,.349,.1}

\journal{Results in Physics}

\begin{document}

\verso{C. A. Wilson-Hodge \textit{etal}}

\begin{frontmatter}

\title{STROBE-X: X-ray Timing and Spectroscopy on Dynamical Timescales from Microseconds to Years}%

\author[1]{Colleen A. \snm{Wilson-Hodge}\corref{cor1}}
\cortext[cor1]{Corresponding author: 
 NASA/MSFC/ST12, 320 Sparkman Dr., Huntsville, AL 35805, USA}
 \ead{colleen.wilson@nasa.gov}
\address[1]{NASA/Marshall Space Flight Center, Huntsville, AL, USA}
\author[2]{Paul S. \snm{Ray}}
\address[2]{Naval Research Lab, Washington, DC, USA}
\author[address3]{Keith \snm{Gendreau}}
\address[address3]{NASA/Goddard Space Flight Center, Greenbelt, MD, USA}
\author[address4]{Deepto \snm{Chakrabarty}}
\address[address4]{MIT Kavli Institute for Astrophysics and Space Research, Cambridge, MA, USA}
\author[address5]{Marco \snm{Feroci}}
\address[address5]{INAF-IAPS, Rome, Italy}
\author[address3,altaddress1]{Zaven \snm{Arzoumanian}}
\address[altaddress1]{USRA, Columbia, MD}
\author[address10]{Soren \snm{Brandt}}
\address[address10]{Technical University of Denmark, Denmark}
\author[address11]{Margarita \snm{Hernanz}}
\address[address11]{Institute of Space Sciences, CSIC-IEEC, Barcelona, Spain}
\author[1]{C.Michelle \snm{Hui}}
\author[address7]{Peter A. \snm{Jenke}}
\address[address7]{University of  Alabama in Huntsville, Huntsville, AL}
\author[address8]{Thomas \snm{Maccarone}}
\address[address8]{Texas Tech University, Lubbock, TX, USA}
\author[address4]{Ron \snm{Remillard}}
\author[address2,altaddress2]{Kent \snm{Wood}}
\address[altaddress2]{Praxis, Inc.}
\author[address9]{Silvia \snm{Zane}}
\address[address9]{Mullard Space Science Laboratory, University College London, UK}
\author{for the STROBE-X collaboration}

\received{01 xxx 2017}
\accepted{01 xxx 2017}
\availableonline{01 xx 2017}

\begin{abstract}
The Spectroscopic Time-Resolving Observatory for Broadband Energy X-rays (STROBE-X) probes 
strong gravity for stellar mass to supermassive black holes and ultradense matter with unprecedented effective area, high time-resolution, and good spectral resolution, while providing a powerful time-domain X-ray observatory.
\end{abstract}


\end{frontmatter}

\section{Introduction}
\label{sec1}
The high-energy sky is extremely dynamic, requiring both wide-field monitoring, to catch a source at the right time, and highly flexible scheduling, to quickly repoint for detailed studies of critical events. Studies of strong gravity and ultradense matter require large collecting areas with low detector deadtime to access the shortest timescales. Broad energy coverage with good spectral resolution is needed to accurately determine continuum spectral shape, to characterize spectral features such as iron lines, to constrain absorption, and to accurately measure the relationship between thermal and non-thermal components. A flexible, high-throughput observatory, the Spectroscopic Time-Resolving Observatory for Broadband Energy X-rays (STROBE-X) has been selected as one of NASA's Astrophysics Probes Mission Concept Studies. These studies will provide input to the 2020 Astrophysics Decadal Survey. STROBE-X serves a large community in a decade of multi-wavelength time-domain astronomy with unique and complementary capabilities to the large high spectral and spatial resolution missions.

\section{Science}
STROBE-X's key science goals include:
\begin{itemize}
\item{Probing stationary spacetimes near black holes (BHs) to explore the effects of strong-field general relativity and measure the masses and spins of BHs, using multiple techniques that allow for cross-calibration.}
\item{X-ray reverberation mapping of the geometry of BH accretion flows across all mass scales, from stellar-mass BHs in our Galaxy to supermassive BHs in active galactic nuclei. }
\item{Fully determining the ultradense matter equation of state by measuring the neutron star mass-radius relation using a large number of pulsars with multiple methods to mitigate systematic uncertainties over an extended mass range.
}
\item{Exploring cosmic chemical evolution by measuring bulk metallicity for numerous high-redshift ($z>2$) clusters.}
\item{Continuously surveying the dynamic X-ray sky with a large duty cycle plus high spectral and time resolution to characterize source behavior over a vast range of time scales. This enables multi-messenger and multi-wavelength studies through cross-correlation with time-domain observatories such as LIGO/Virgo, IceCube, LSST, and SKA.
} 
\end{itemize}

\begin{figure}
\centering
\includegraphics[width=3.25in]{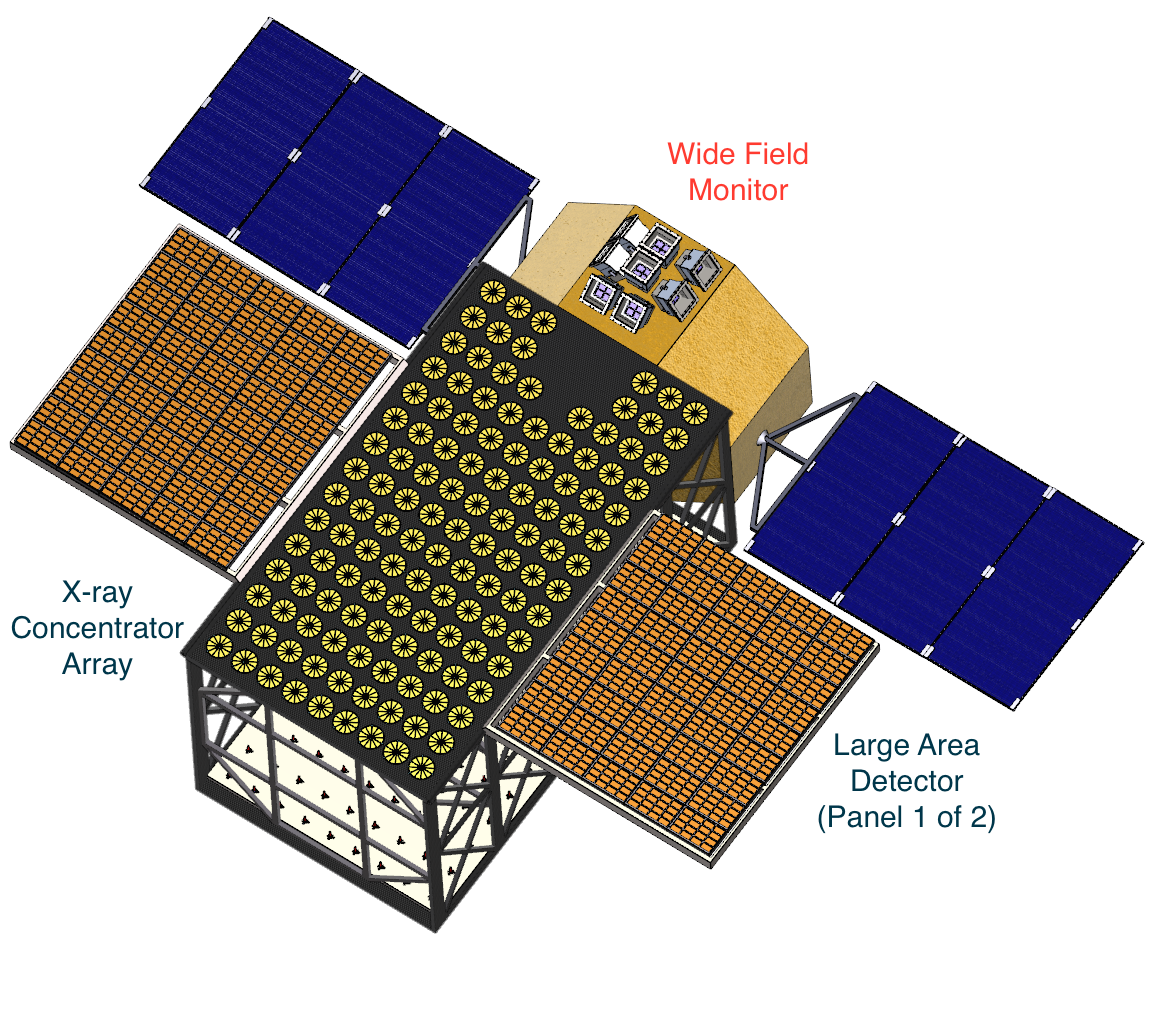} 
\vskip -0.13 in
\caption{Notional deployed configuration of the STROBE-X spacecraft.\label{fig-strobex}}
\vskip -0.1 in
\end{figure}

\section{Mission Concept}

STROBE-X is planned for a Falcon 9 launch into an orbit with as low an inclination as possible. The satellite 
bus and mission operations are designed to allow rapid ($\sim$hours) and autonomous ($\sim$minutes) repointing.  STROBE-X 
comprises three instruments as shown in Figure~\ref{fig-strobex}. The soft band (0.2--12 keV) is covered by 
the X-ray Concentrator Array (XRCA), an array of lightweight optics (3-m focal length) that concentrate 
incident photons onto small solid state detectors with CCD-like (85--130 eV) energy resolution, 100 ns time 
resolution, and low background rates. The harder band (2 to at least 30 keV) is covered by the Large Area 
Detector (LAD,\cite{Zane14}), comprising large-area silicon drift detectors (SDDs), with 200--240 eV energy resolution, collimated 
to a 1$^\circ$ field-of-view with lead-glass micropore collimators. Each instrument would provide an order of 
magnitude improvement in effective area compared with its predecessor (NICER in the soft band and RXTE in the 
hard band). A sensitive wide-field sky monitor (WFM,\cite{Brandt14}) would act as a trigger for pointed observations, provide high duty cycle, 
high time resolution, high spectral resolution monitoring of the X-ray sky with $\sim$10 times the sensitivity of 
the RXTE All-Sky Monitor \cite{Levine96}, and enable multi-wavelength and multi-messenger studies on a 
continuous, rather than scanning basis. Continuous telemetry of the WFM data will make it a powerful instrument in its own right.

STROBE-X builds upon the X-ray timing results, existing technologies, and community built from the Rossi 
X-ray Timing Explorer (RXTE, 1995--2012, \cite{Swank06}), the Large Observatory For x-ray Timing (LOFT, 
\cite{Feroci12, Feroci14}), studies for the Advanced X-ray Timing Array (AXTAR, \cite{Ray10}) and LOFT-Probe (LOFT-P, \cite{WilsonHodge16}), and the Neutron star Interior Composition Explorer 
(NICER, 2017--present, \cite{Gendreau16}). The X-ray concentrator optics, fully developed for NICER, are scaled up with longer focal-lengths to provide large collecting area with low background at low cost. SDDs, developed for 
LOFT, provide high time resolution with low dead time and CCD-like spectroscopy.  Micropore collimators have 
dramatically less mass and volume than traditional designs, enabling large missions at modest cost. 

\begin{figure}
\centering
\includegraphics[width=3.25in]{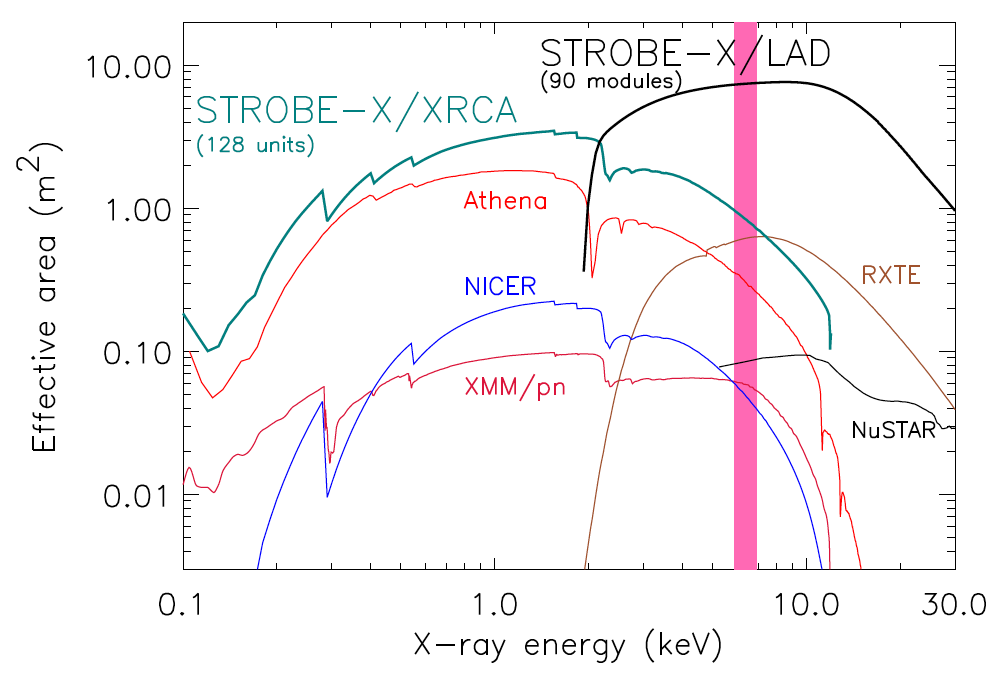}
\vskip -0.13 in
\caption{Effective area of the two STROBE-X pointed instruments (XRCA and LAD) for the baseline configuration shown in Figure~\ref{fig-strobex} with 90 LAD modules and 128 XRC
units. The critical iron line region is marked with the red bar. \label{fig-effarea} }
\vskip -0.1 in
\end{figure}





\bibliographystyle{model3-num-names}
\bibliography{refs}

\begin{thebibliography}{9}
\providecommand{\natexlab}[1]{#1}
\providecommand{\url}[1]{\texttt{#1}}
\providecommand{\href}[2]{#2}
\providecommand{\path}[1]{#1}
\providecommand{\eprint}[1]{\href{http://arxiv.org/abs/#1}{\path{#1}}}
\providecommand{\DOIprefix}{doi:}
\providecommand{\ArXivprefix}{arXiv:}
\providecommand{\URLprefix}{URL: }
\providecommand{\Pubmedprefix}{pmid:}
\providecommand{\doi}[1]{\href{http://dx.doi.org/#1}{\path{#1}}}
\providecommand{\Pubmed}[1]{\href{pmid:#1}{\path{#1}}}
\providecommand{\BIBand}{and}
\providecommand{\bibinfo}[2]{#2}
\ifx\xfnm\undefined \def\xfnm[#1]{\unskip,\space#1}\fi
\bibitem[{{Zane} et~al.(2014)}]{Zane14}
\bibinfo{author}{{Zane}\xfnm[ S.]}, et~al.
\newblock \bibinfo{title}{{The large area detector of LOFT: the Large
  Observatory for X-ray Timing}}.
\newblock In: \bibinfo{booktitle}{Space Telescopes and Instrumentation 2014:
  Ultraviolet to Gamma Ray}; vol. \bibinfo{volume}{9144} of
  \emph{\bibinfo{series}{Proc. SPIE}}. \bibinfo{year}{2014}, p.
  \bibinfo{pages}{91442W}.
\bibitem[{{Brandt} et~al.(2014)}]{Brandt14}
\bibinfo{author}{{Brandt}\xfnm[ S.]}, et~al.
\newblock \bibinfo{title}{{The design of the wide field monitor for the LOFT
  mission}}.
\newblock In: \bibinfo{booktitle}{Space Telescopes and Instrumentation 2014:
  Ultraviolet to Gamma Ray}; vol. \bibinfo{volume}{9144} of
  \emph{\bibinfo{series}{Proc. SPIE}}. \bibinfo{year}{2014}, p.
  \bibinfo{pages}{91442V}.
\bibitem[{{Levine} et~al.(1996)}]{Levine96}
\bibinfo{author}{{Levine}\xfnm[ A.M.]}, et~al.
\newblock \bibinfo{title}{{First Results from the All-Sky Monitor on the Rossi
  X-Ray Timing Explorer}}.
\newblock \bibinfo{journal}{ApJ}
  \bibinfo{year}{1996};\bibinfo{volume}{469}:\bibinfo{pages}{L33}.
\bibitem[{{Swank}(2006)}]{Swank06}
\bibinfo{author}{{Swank}\xfnm[ J.H.]}.
\newblock \bibinfo{title}{{The Rossi X-ray timing explorer: Capabilities,
  achievements and aims}}.
\newblock \bibinfo{journal}{AdSpR}
  \bibinfo{year}{2006};\bibinfo{volume}{38}:\bibinfo{pages}{2959--2963}.
\bibitem[{{Feroci} et~al.(2012)}]{Feroci12}
\bibinfo{author}{{Feroci}\xfnm[ M.]}, et~al.
\newblock \bibinfo{title}{{The Large Observatory for X-ray Timing (LOFT)}}.
\newblock \bibinfo{journal}{ExA}
  \bibinfo{year}{2012};\bibinfo{volume}{34}:\bibinfo{pages}{415--444}.
\bibitem[{{Feroci} et~al.(2014)}]{Feroci14}
\bibinfo{author}{{Feroci}\xfnm[ M.]}, et~al.
\newblock \bibinfo{title}{{The Large Observatory for x-ray timing}}.
\newblock In: \bibinfo{booktitle}{Space Telescopes and Instrumentation 2014:
  Ultraviolet to Gamma Ray}; vol. \bibinfo{volume}{9144} of
  \emph{\bibinfo{series}{Proc. SPIE}}. \bibinfo{year}{2014}, p.
  \bibinfo{pages}{91442T}.
\bibitem[{{Ray} et~al.(2010)}]{Ray10}
\bibinfo{author}{{Ray}\xfnm[ P.S.]}, et~al.
\newblock \bibinfo{title}{{AXTAR: mission design concept}}.
\newblock In: \bibinfo{booktitle}{Space Telescopes and Instrumentation 2010:
  Ultraviolet to Gamma Ray}; vol. \bibinfo{volume}{7732} of
  \emph{\bibinfo{series}{Proc. SPIE}}. \bibinfo{year}{2010}, p.
  \bibinfo{pages}{773248}.
\bibitem[{{Wilson-Hodge} et~al.(2016)}]{WilsonHodge16}
\bibinfo{author}{{Wilson-Hodge}\xfnm[ C.A.]}, et~al.
\newblock \bibinfo{title}{{Large Observatory for x-ray Timing (LOFT-P): a
  Probe-class mission concept study}}.
\newblock In: \bibinfo{booktitle}{Society of Photo-Optical Instrumentation
  Engineers (SPIE) Conference Series}; vol. \bibinfo{volume}{9905} of
  \emph{\bibinfo{series}{Proc. SPIE}}. \bibinfo{year}{2016}, p.
  \bibinfo{pages}{99054Y}.
\bibitem[{{Gendreau} et~al.(2016)}]{Gendreau16}
\bibinfo{author}{{Gendreau}\xfnm[ K.C.]}, et~al.
\newblock \bibinfo{title}{{The Neutron star Interior Composition Explorer
  (NICER): design and development}}.
\newblock In: \bibinfo{booktitle}{Society of Photo-Optical Instrumentation
  Engineers (SPIE) Conference Series}; vol. \bibinfo{volume}{9905} of
  \emph{\bibinfo{series}{Proc. SPIE}}. \bibinfo{year}{2016}, p.
  \bibinfo{pages}{99051H}.

\end{thebibliography}

\end{document}